\documentclass[pra, showpacs, twocolumn, floatfix,superscriptaddress]{revtex4-2}
\usepackage{graphicx}
\usepackage{amsmath, amsfonts, amssymb, bm, ulem, color}

\begin{document}
	%%%%%%%%%%%%%%%%%%%%%%%%%%%%%%%%%%%%%%%%%%%%%%%%%%%%%%%%%%%%%%%%%
	\title{Enhanced phonon lifetimes with optically controlled single molecules}
	\author{Victor \surname{Ceban}}
	\email{victor.ceban@ifa.md}
	\affiliation{Institute of Applied Physics, Moldova State University,
		Academiei str. 5, MD-2028 Chi\c{s}in\u{a}u, Moldova}
	\author{Mihai A. \surname{Macovei}}
	\email{mihai.macovei@ifa.md}
	\affiliation{Institute of Applied Physics, Moldova State University,
		Academiei str. 5, MD-2028 Chi\c{s}in\u{a}u, Moldova}

	%%%%%%%%%%%%%%%%%%%%%%%%%%%%%%%%%%%%%%%%%%%%%%%%%%%%%%%%%%%%%%%%%
\begin{abstract}
We have investigated the phonon dynamics of a single-molecule embedded in a mechanical resonator made of an organic crystal.
The whole system is placed in an optical resonator within the bad cavity limit.
We have found that the optical control of the molecular population affects the phonon dynamics. Long-lived phonons are obtained when slowing-down the decay dynamics of the molecule via modulation of the transition frequency. The discussed results are also valid for optomechanical setups based on other types of two-level emitters and mechanical resonators.
\end{abstract}
%%%%%%%%%%%%%%%%%%%%%%%%%%%%%%%%%%%%%%%%%%%%%%%%%%%%%%%%%%%%%%%%%%%
\maketitle
%%%%%%%%%%%%%%%%%%%%%%%%%%%%%%%%%%%%%%%%%%%%%%%%%%%%%%%%%%%%%%%%%%%
%%%%%%%%%%%%%%%%%%%%%%%%%%%%%%%%%%%%%%%%%%%%%%%%%%%%%%%%%%%%%%%%%%%
\section{Introduction}
%%%%%%%%%%%%%%%%%%%%%%%%%%%%%%%%%%%%%%%%%%%%%%%%%%%%%%%%%%%%%%%%%%%

During the last decade, significant advancements have been achieved in experimental and technological applications of quantum optics. The broad gamut of achievements varies from the computing, transfer and storage of quantum information to the construction of on-chip devices able to reproduce various quantum effects on-demand. This progress has strengthened the demand for more sophisticated and tunable quantum emitters \cite{sart22, mort21}. In this context, molecules represent a promising candidate due to their rich and various energetic structures \cite{toni21}.
Remarkable observations of two-photon interferences \cite{treb10, kira05} and quantum entaglament of two independent photons \cite{reza19} emitted from a single molecule have been recently reported. Single-molecule-based optical transistors \cite{hwan09}, quantum gates \cite{kewe16}, optomechanical detectors of vibrational quantum motion \cite{kolc09}, i.e., nanomicrophones, as well as, pressure sensing technique \cite{croc93} had been demonstrated.
Polar molecules are expected to serve as quantum memory with large coherence time when collectively interacting within an optical cavity with a superconducting circuit in it \cite{rabl06}. Phonon-assisted single-photon up-conversion schemes from mid-infra-red to visible domain with molecules had been predicted in \cite{roel20}.
Moreover, strong coupling regimes had been already observed at room termperatures with methylene-blue molecules placed in optical nanocavities \cite{chik16}. This kind of metal optical cavities of few nanometers in size allows the confinement of optical fields with much larger wavelengths \cite{benz16}, while the surface plasmons may exhibit strong optomechanical couplings with the vibrations of the embedded molecule \cite{hugh21}.

Particularly promising quantum emitters, with a high degree of coherence, are obtained when operating polycyclic aromatic hydrocarbon molecule at cryogenic temperatures of a few Kelvins.
At such temperature, the molecular energetic bandwidths  become narrow enough so that the molecule can be used as a two-level emitter. 
An optical coupling of $g \approx 0.8 \rm{GHz}$ had been experimentally reported when placing such a molecule in a microcavity \cite{wang19}. These emitters had also shown excellent first-order coherence in the generation of single-photons \cite{reza18}. 
Such phenomena as Mollow triplet \cite{wrig08}, photon antibunching and Rabi oscillations \cite{basc92} have been recently observed with aromatic molecules.
Moreover, an on-chip realisation of this type of emitters coupled to a waveguide had been reported in \cite{lomb18}, with a coupling efficiency of up to $\approx 42\%$.

When the polycyclic aromatic hydrocarbon molecule is embedded in a solid organic crystal,
an optomechanical coupling with the crystal phonons occurs. The phonon state density distribution of the solid changes when the dimensions are reduced to sizes of tens or hundreds of nanometers. 
At this sizes, various crystal geometries are expected to drastically reduce the phonon state distribution to the limit of single-mode phonon fields \cite{aspe14}. A recent observation of the quantum dynamics of optomechanical interaction of an aromatic molecule with single-mode phonons of an organic crystal had been reported \cite{gurl21}. Long-lived phonons had been observed by reducing the size of the organic crystal and by engineering the geometry of its substrate. A further decrease of the phonon damping rate, would allow the storage of optical information in an optomechanical device for durations of milliseconds \cite{gurl21}. Such hybrid quantum memory would be competitive with state-of-the-art optomechanical quantum memories based on whispering gallery mode mechanical resonators \cite{lake21}, membrane mechanical resonators \cite{liu23} and Brillouin scattering in waveguides \cite{merk17}.

 %%%%%%%%%%%%%%%%%%%%%%%%%%%%%%%%%%%%%%%%%%%%%%%%%%%%%%%%%%%%%%
\begin{figure}[b]
	\centering
	\includegraphics[width= 7.5cm ]{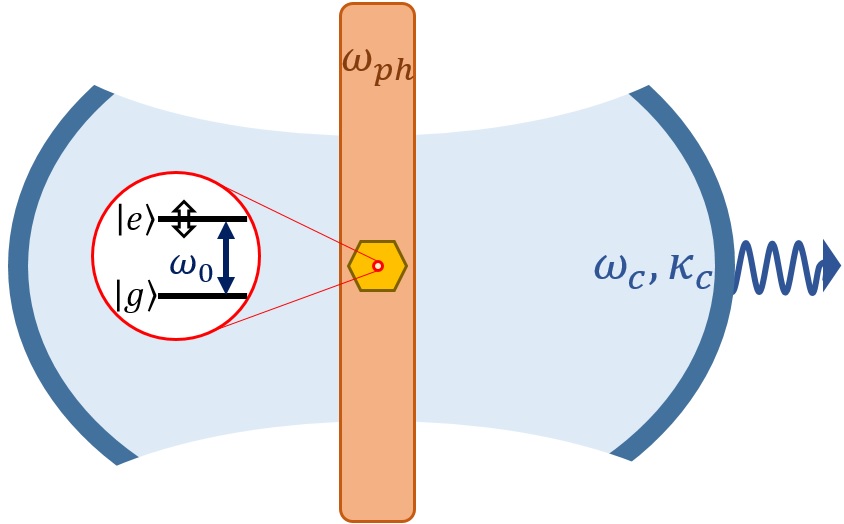}
	\caption{\label{model} 
		The schematic of the model: An aromatic molecule is embedded within an organic crystal placed on a mechanical resonator. The optomechanical system is placed in an optical cavity.}
\end{figure}
%%%%%%%%%%%%%%%%%%%%%%%%%%%%%%%%%%%%%%%%%%%%%%%%%%%%%%%%%%%%%%

In this paper, we investigate the possibility to enhance phonon lifetimes in an optomechanical system made of a polycyclic aromatic hydrocarbon molecule embedded within an organic crystal placed on a mechanical resonator.
The molecule acts as a two-level emitter and its spontaneous emission is slowed-down.
Optical control over the spontaneous emission is achieved by modulating the molecular transition frequency and placing the molecule within an optical cavity with a low quality factor \cite{maco14}, as shown in Fig.\ref{model}.
The frequency modulation may be obtained via dynamic Stark effect. In this regard, large linear Stark shifts had been reported for aromatic molecules \cite{mora19}.
The slowed-down decay dynamics results in slowing-down the phonon dynamics of the mechanical resonator.
This is oppositely to fast phonon dynamics occurring due to superradiance effects in related setups \cite{ceba17, mori19}.

It is to note that the investigated quantum model can be also applied for other types of optomechanical devices, e.g., quantum dots placed on mechanical resonators or nitrogen vacancy in diamonds. Moreover, a similar behaviour of slowed-down decay dynamics may be achieved by other techniques such as fast modulation of the atomic transition and its system-bath coupling \cite{agar99, jano00} or modulation of reservoir modes \cite{lini08}. However, the chosen model can be also applied in the framework of molecular optomechanics in order to describe the resonant interaction of a plasmonic field hot-spot of a metal nanoparticle with the vibrations of a single-molecule \cite{hugh21}. The observation of Stark shifts and Purcell effect in such systems had been reported in \cite{rosl22}.  

This paper is organized as follows. In Sec. II, we present the analytic model and describe the solving techniques of the quantum dynamic equations. In Sec. III, we discuss the obtained results, while in Sec. IV we present our conclusions.

%%%%%%%%%%%%%%%%%%%%%%%%%%%%%%%%%%%%%%%%%%%%%%%%%%%%%%%%%%%%%%%%%%%
\section{The Model }
%%%%%%%%%%%%%%%%%%%%%%%%%%%%%%%%%%%%%%%%%%%%%%%%%%%%%%%%%%%%%%%%%%%
The molecule is described as a two-level emitter with the excited state $\vert e \rangle$ and ground state $\vert g \rangle$. The corresponding atomic operators are defined as: $S^{+}= \vert e \rangle \langle g \vert$, $S^{-}= \vert g \rangle \langle e \vert$, $S_{ee}=\vert e \rangle \langle e \vert$ and $S_z= (\vert e \rangle \langle e \vert-\vert g \rangle \langle g \vert)/2$. They obey the standard commutation relations for SU(2) algebra, i.e., $[S^\pm,S^\mp]=\pm 2 S_z$ and $[S^\mp,S_z ]=\pm S^\mp$. The modulated transition frequency of the molecule is composed by the free transition frequency $\omega_0$ and the modulation signal $s(t)$. The optical cavity is defined by the cavity frequency $\omega_{c}$, the cavity damping rate $\kappa_c$ and the photon annihilation and creation operators $a$ and $a^{\dagger}$, respectively. The single-mode phonon field of the quantum mechanical resonator is described by its phonon frequency $\omega_{ph}$ and its annihilation and creation operators $b$ and $b^{\dagger}$, respectively. The damping of the mechanical mode by the thermal environment is defined by the damping rate of the mechanical resonator $\kappa$ and the mean phonon number of the thermal bath $\bar{n} = [\exp (\hbar \omega_{ph} / k_{B}T) -1]^{-1}$, where $T$ is the environmental temperature and $k_{B}$ is the Boltzmann constant.

Thus, the system Hamiltonian is given as:
%%%%%%%%%%%%%%%%%%%%%%%%%%%%%%%%%%%%%%%%%%%%%%%%%%%%%%
\begin{eqnarray}
	H &=& \hbar \omega_{c} a^{\dagger}a + \hbar \omega_{ph} b^{\dagger}b + \hbar \left[ \omega_{0} + s(t) \right] S_{z}  \nonumber \\
	 &+&  \hbar g \left(a^{\dagger}S^{-}+ S^{+}a \right)+ \hbar \lambda S_{ee} \left(b+b^{\dagger} \right).
	\label{Hinit}
\end{eqnarray}
%%%%%%%%%%%%%%%%%%%%%%%%%%%%%%%%%%%%%%%%%%%%%%%%%%%%%%
Here, the first term describes the free Hamiltonian of the optical cavity. The second term defines the free mechanical resonator. The third term describes the free molecule with a modulated transition frequency. The fourth term describes the quantum interaction of the molecule with the optical cavity and is defined by the coupling constant $g$. The last term describes the interaction of the molecule with the phonons of the mechanical resonator and is defined by the coupling constant $\lambda$. The couplings $g$ and $\lambda$ are considered constant for weak modulations of the transition frequency, i.e., $s(t) \ll \omega_0$.

The system dynamics is expressed by the master equation of the density operator $\rho$, given as:
%%%%%%%%%%%%%%%%%%%%%%%%%%%%%%%%%%%%%%%%%%%%%%%%%%%%%%
\begin{eqnarray}
	\frac{\partial \rho}{\partial t} &=& -\frac{i}{\hbar} \left[H,\rho\right]+\kappa(1+\bar{n}) \mathcal{L}(b)                         
	+\kappa\bar{n}\mathcal{L}(b^{\dagger})    \nonumber  \\
	&+& \kappa_c \mathcal{L}(a)	,
	\label{meq}
\end{eqnarray}
%%%%%%%%%%%%%%%%%%%%%%%%%%%%%%%%%%%%%%%%%%%%%%%%%%%%%%
where the Liouville superoperator $\mathcal{L}$ is defined for a given operator $\mathcal{O}$ as: $ \mathcal{L}(\mathcal{O})= 2\mathcal{O}\rho \mathcal{O}^{\dagger}-\mathcal{O}^{\dagger} \mathcal{O}\rho-\rho \mathcal{O}^{\dagger} \mathcal{O} $. Here, the first term represents the von Neumann equation of the system Hamiltonian $H$. The second and the third terms represent the phonon leaking and pumping terms, respectively, which appear due to the interaction of the mechanical resonator with the thermal bath. The last term describes the damping of the optical cavity.

The first step in solving the system dynamics consists in eliminating the photonic operators. This is done by considering a periodic modulation signal which can be decomposed in Fourier series as: $s(t)= \sum_{0}^{\infty} A_j \cos(j \omega t +\phi_j)$, where for the sake of simplicity we consider $A_0=0$ and $\phi_j=0$. For such type of signals, the system Hamiltonian $H$ can be expressed in a new frame defined by the unitary transformation $U(t)= \exp[-i/h \int^{t}_{0}H_0 (\tau) d \tau ]$ where $H_0(t)= \hbar \omega_{0} a^{\dagger}a + \hbar \left[ \omega_{0} + s(t) \right] S_{z}$. Within this frame, $H$ is expressed as:
%%%%%%%%%%%%%%%%%%%%%%%%%%%%%%%%%%%%%%%%%%%%%%%%%%%%%%
\begin{eqnarray}
	H &=& i g \prod_{j=1}^{\infty} \sum_{m=-\infty}^{\infty} J_m \left(\frac{A_j}{j\omega} \right) \left[S^{-} a^{\dagger} e^{-ijm\omega t}-a S^{+} e^{ijm \omega t} \right] \nonumber \\                                   
	&+&  \hbar \delta_c a^{\dagger}a + \hbar \omega_{ph} b^{\dagger} b + \hbar \lambda S_{ee}(b^{\dagger}+b) ,          	  	
	\label{Hfull}
\end{eqnarray}
%%%%%%%%%%%%%%%%%%%%%%%%%%%%%%%%%%%%%%%%%%%%%%%%%%%%%%
where $\delta_c=\omega_{c}-\omega_{0}$. Here, one have applied the Jacobi-Anger expansion: $e^{i A \sin(\omega t + \phi)}= \sum_{-\infty}^{\infty} J_m (A) e^{im(\omega t + \phi)}$ where $J_m$ are the Bessel functions of the first kind.

The infinite product of series eq.(\ref{Hfull}) is further simplified as follows:
\begin{eqnarray}
	&& \prod_{j=1}^{\infty} \sum_{m=-\infty}^{\infty}J_m  \left(\frac{A_j}{j\omega} \right)  \left[S^{-} a^{\dagger} e^{-ijm\omega t}-a S^{+} e^{ijm \omega t} \right]  \nonumber \\ 
	&&  = \sum_{m=-\infty}^{\infty} C_{m,\infty}  \left[S^{-} a^{\dagger} e^{-im\omega t}-a S^{+} e^{im \omega t} \right] .
	\label{coeff}
\end{eqnarray}
The coefficients $C_{m,j}$ are obtained from the $j^{th}$ multiplication of two succesive power series and are expressed as:  $C_{m,j}= \sum_{k=-\infty}^{m} C_{k,j-1} J_{m-k} (A_j / j\omega) \delta_{(m-k)/j,[(m-k)/j]} $ where $C_{m,1}=J_m \left(\frac{A_1}{\omega} \right)$, $\delta_{x,[x]}$ is the Kronecker function and $[x]$ is the integer part of a number $x$.

By considering the simplified form of the Hamiltonian, it is possible to identify and solve the equations of motion of the photonic operators of the optical cavity within the bad cavity limit, i.e., for $ \kappa_c \gg g$, and for the Born-Markov approximations, similarly to the method applied in \cite{maco14}.

The obtained expression of the creation operator $a^{\dagger}(t)$:  
%%%%%%%%%%%%%%%%%%%%%%%%%%%%%%%%%%%%%%%%%%%%%%%%%%%%%%
\begin{eqnarray}
a^{\dagger} (t)&=& a^{\dagger} (0) e^{-(\kappa_c-i\delta _c )t} \nonumber \\ 
&&+g\sum_{m=-\infty}^{\infty} \frac{C_{m,\infty} e^{im\omega t}}{\kappa_c+i(m\omega-\delta _c )}  S^{+} ,  
\end{eqnarray}
%%%%%%%%%%%%%%%%%%%%%%%%%%%%%%%%%%%%%%%%%%%%%%%%%%%%%%
and its Hermitian conjugate $a(t)$ are inserted within the master equation of eq.(\ref{meq}) expressed within the Heisenberg representation, in order to eliminate the photonic operators from the equation. From the new form of the master equation, the equation of motion of a given operator of the molecule-mechanical-resonator sub-system $Q$ is obtained as follows:
%%%%%%%%%%%%%%%%%%%%%%%%%%%%%%%%%%%%%%%%%%%%%%%%%%%%%%
\begin{eqnarray}
	\frac{d \left\langle  Q\right\rangle}{dt}  &=&  i\omega_{ph} \left\langle   [b^{\dagger} b,Q]\right\rangle + i\lambda\left\langle [ (b^{\dagger}+b)S_{ee},Q]  \right\rangle   \nonumber \\
	&+&i\Omega(t)\left\langle  [S_z,Q]\right\rangle +\gamma(t)\mathcal{L}'(S^{-})          \nonumber \\
	&+&\kappa(1+\bar{n}) \mathcal{L}'(b)                         
	+\kappa\bar{n}\mathcal{L}'(b^{\dagger}) , 
	\label{meqfinal} 
\end{eqnarray}
%%%%%%%%%%%%%%%%%%%%%%%%%%%%%%%%%%%%%%%%%%%%%%%%%%%%%%
where $ \mathcal{L}'(\mathcal{O})= 2\left\langle \mathcal{O}Q \mathcal{O}^{\dagger}\right\rangle -\left\langle \mathcal{O}^{\dagger} \mathcal{O}Q\right\rangle  - \left\langle Q \mathcal{O}^{\dagger} \mathcal{O}\right\rangle   $. The pseudo-damping coefficients which appear after eliminating the photonic operators are defined as:
%%%%%%%%%%%%%%%%%%%%%%%%%%%%%%%%%%%%%%%%%%%%%%%%%%%%%%
\begin{eqnarray}
\gamma(t) &=& \sum_{\{m,n=-\infty\} }^{\infty} C_{m,\infty} C_{n,\infty}   \nonumber \\
&&\quad \times \frac{g^2 \kappa_c }{\kappa_c^2 + (m\omega-\delta _c )^2}   \cos{\left( (m-n)\omega t \right) }  ,  \nonumber \\
\Omega(t) &=& \sum_{\{m,n=-\infty\} }^{\infty} C_{m,\infty} C_{n,\infty}  \nonumber \\
&& \quad \times  \frac{g^2 (m\omega -\delta_c) }{\kappa_c^2 + (m\omega-\delta _c )^2}   \cos{\left( (m-n)\omega t \right) } .
\label{coeffs}  
\end{eqnarray}
%%%%%%%%%%%%%%%%%%%%%%%%%%%%%%%%%%%%%%%%%%%%%%%%%%%%%%

The system dynamics is determined from the equations of motion of the parameters of interest, built from the general form of the equation of motion of eq.(\ref{meqfinal}). Namely, the dynamics of the molecule excited state population $\left\langle  S_{ee} \right\rangle$ is described by:
%%%%%%%%%%%%%%%%%%%%%%%%%%%%%%%%%%%%%%%%%%%%%%%%%%%%%%
\begin{eqnarray}
\frac{d \left\langle  S_{ee} \right\rangle}{dt}  &=&- 2 \gamma(t) \left\langle  S_{ee} \right\rangle . 
\label{eqs22}   
\end{eqnarray}
%%%%%%%%%%%%%%%%%%%%%%%%%%%%%%%%%%%%%%%%%%%%%%%%%%%%%%

The mean phonon number of the quantum mechanical resonator $\left\langle  b^{\dagger} b\right\rangle$ is deduced from the following system: 
%%%%%%%%%%%%%%%%%%%%%%%%%%%%%%%%%%%%%%%%%%%%%%%%%%%%%%
\begin{eqnarray}
\frac{d \left\langle  b^{\dagger} b\right\rangle}{dt}  &=&i\lambda\{\left\langle  S_{ee} b\right\rangle  -\left\langle  S_{ee} b^{\dagger} \right\rangle    \}-2\kappa\left\langle  b^{\dagger} b\right\rangle  +2\kappa\bar{n} , \nonumber \\
\frac{d\left\langle  S_{ee} b\right\rangle}{dt}  &=&-(2\gamma(t)+i\omega_{ph} +\kappa)\left\langle  S_{ee} b\right\rangle   -i \lambda \left\langle  S_{ee} \right\rangle  .  
\label{eqbb}
\end{eqnarray}
%%%%%%%%%%%%%%%%%%%%%%%%%%%%%%%%%%%%%%%%%%%%%%%%%%%%%%
The missing equation of motion of $\left\langle  S_{ee} b^{\dagger}\right\rangle$ is obtained by Hermitian conjugation of the equation of motion of $\left\langle  S_{ee} b\right\rangle$.

The dynamics of the second-order phonon-phonon correlation function $g^{(2)}(0)=\left\langle  b^{\dagger 2} b^{2}\right\rangle / \left\langle  b^{\dagger} b\right\rangle ^2$ is obtained from the following system of equations:
%%%%%%%%%%%%%%%%%%%%%%%%%%%%%%%%%%%%%%%%%%%%%%%%%%%%%%
\begin{eqnarray}
\frac{d \left\langle  b^{\dagger 2} b^{2}\right\rangle }{dt}  &=& 2i\lambda \left\lbrace  \left\langle  S_{ee} b^{\dagger} b^{2}\right\rangle - \left\langle  S_{ee} b^{\dagger 2}  b\right\rangle \right\rbrace \nonumber \\
&-&4\kappa\left\langle  b^{\dagger 2} b^{2} \right\rangle  +8\kappa\bar{n}\left\langle  b^{\dagger} b\right\rangle ,  \nonumber \\
\frac{d\left\langle  S_{ee} b^{\dagger}b^{2} \right\rangle }{dt} &=&-(2\gamma(t)+i\omega_{ph} +3\kappa)\left\langle S_{ee} b^{\dagger}b^{2} \right\rangle  \nonumber\\
&+& i\lambda \left\lbrace  \left\langle  S_{ee} b^{2} \right\rangle -2 \left\langle  S_{ee} b^{\dagger}b \right\rangle  \right\rbrace  +4\kappa\bar{n}\left\langle  S_{ee} b\right\rangle ,   \nonumber \\
\frac{d\left\langle  S_{ee} b^{2}\right\rangle }{dt} &=&-2(\gamma(t)+i\omega_{ph} +\kappa)\left\langle  S_{ee} b^{2}\right\rangle  -2i\lambda\left\langle  S_{ee} b\right\rangle ,  \nonumber \\
\frac{d\left\langle  S_{ee} b^{\dagger} b\right\rangle }{dt} &=&-2(\gamma(t)+\kappa)\left\langle  S_{ee} b^{\dagger} b\right\rangle +2\kappa\bar{n}\left\langle  S_{ee} \right\rangle   \nonumber \\
&+& i\lambda\left\lbrace  \left\langle  S_{ee} b\right\rangle  -\left\langle  S_{ee} b^{\dagger} \right\rangle \right\rbrace    .
\label{eqg2}
\end{eqnarray}
%%%%%%%%%%%%%%%%%%%%%%%%%%%%%%%%%%%%%%%%%%%%%%%%%%%%%%
The missing equations of motion are obtained by Hermitian conjugation of the equations of motion of $\left\langle  S_{ee} b^{\dagger}b^{2} \right\rangle$ and $\left\langle  S_{ee} b^{2}\right\rangle$.

In order to numerically solve the systems of equations of eqs.(\ref{eqs22} - \ref{eqg2}), the infinite series of eqs.(\ref{coeffs}) must be truncated at a certain threshold. This is achieved by considering only the first $j_{max}$ terms of Fourier series used to develop the modulation signal $s(t)$ and the first $m_{max}$  terms of the Jacobi-Anger expansion.
The value of $m_{max}$ is empirically deduced such as further increase of $m_{max}$ does not change the established quantum dynamics. Therefore, the infinite series of eqs.(\ref{coeffs}) are reduced to finite sums limited by $\pm j_{max}^2 m_{max}$, i.e. $\sum_{-\infty}^{\infty} \rightarrow \sum_{-j_{max}^2 m_{max}}^{j_{max}^2 m_{max}}$. Similarly, the multiplication coefficients of eq.(\ref{coeff}) are reduced to $C_{m,\infty} \rightarrow C_{m,j_{max}} $.

\section{Results and discussions}
%%%%%%%%%%%%%%%%%%%%%%%%%%%%%%%%%%%%%%%%%%%%%%%%%

\begin{figure}[b]
 	\centering
 	\includegraphics[width= 7.5cm ]{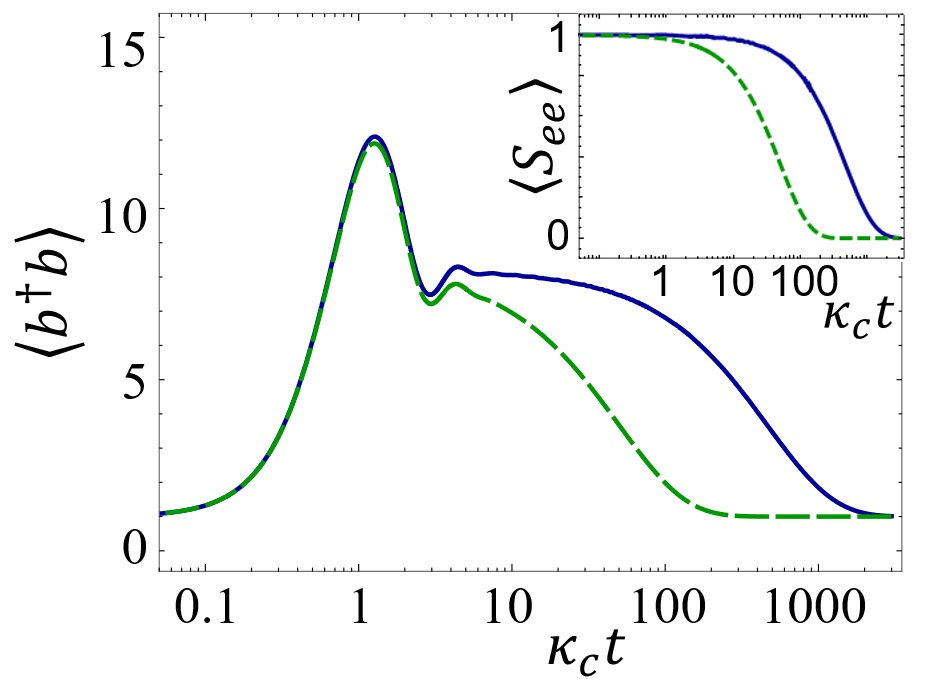}
 	\caption{\label{pop} 
 		Mean phonon number $\left\langle  b^{\dagger} b\right\rangle$ as function of time for non-modulated transition frequency $s(t)=0$ (green dashed line) and for a harmonic modulation of the transition frequency $s(t)=A \cos(\omega t)$ (blue solid line). The inset depicts the molecular excited state population $\left\langle  S_{ee} \right\rangle$ for the same cases. The parameters are given in the text.}
 \end{figure}

\begin{figure}[b]
	\centering
	\includegraphics[width= 7.5cm ]{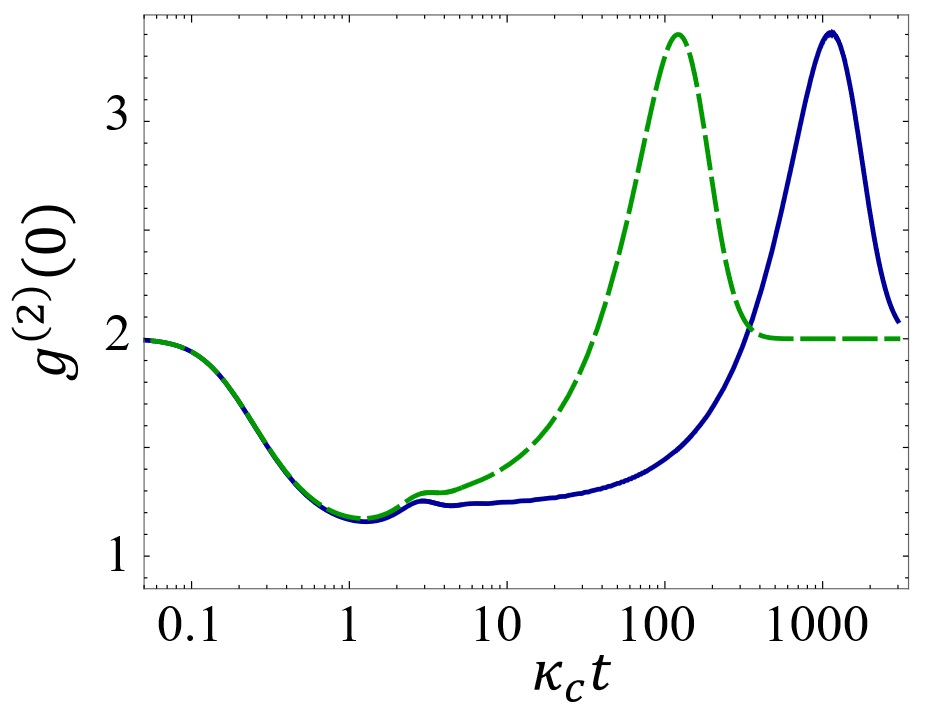}
	\caption{\label{g2} 
		Second-order phonon-phonon correlation function $g^{(2)}(0)$ as function of time for non-modulated transition frequency (green dashed line) and for a harmonic modulation of the transition frequency $s(t)=A \cos(\omega t)$ (blue solid line). The parameters are given in the text.}
\end{figure}

The quantum dynamics of the phonon field is related to the behaviour of the molecular state populations and a slow-down of the molecular population decay affects the phonon lifetimes. In what follows, we investigate the dynamics of the mean phonon number $\left\langle  b^{\dagger} b\right\rangle$ in Fig.\ref{pop} and the second-order phonon-phonon correlation function $g^{(2)}(0)$  in Fig.\ref{g2}, as functions of normalized time $\kappa_c t$, for an initially excited two-level molecule with modulated and non-modulated transition frequencies. We compare these dynamics to the behaviour of the excited state population of the molecule $\left\langle  S_{ee} \right\rangle$ depicted in the inset of Fig.\ref{pop}.
 
 %%%%%%%%%%%%%%%%%%%%%%%%%%%%%%%%%%%%%%%%%%%%%%%%%%%%%%%%%%%%%%
 %%%%%%%%%%%%%%%%%%%%%%%%%%%%%%%%%%%%%%%%%%%%%%%%%%%%%%%%%%%%%%

The case of a molecule with a non-modulated transition frequency is described by $s(t)=0$ and is represented with green dashed lines. The case of a molecule with modulated transition frequency is described by a harmonic modulation signal $s(t)=A \cos(\omega t)$, with $\omega/\kappa_c=1$, $A/\kappa_c=10$, $j_{max}=1$, and is represented with blue solid lines. The bad cavity limit is given by $g/\kappa_c=0.1$. Without loosing on the generality of the problem, the optical cavity in considered in resonance with the molecular transition, i.e., $\omega_{c}=\omega_{0}$. Therefore, the cavity-molecule detuning $\delta_c=0$ and does not affect the system dynamics. The molecule is initially prepared in an excited state, e.g., by a short laser pulse. We assume that the preparation phase does not affect the system dynamics described in the previous section, as long as the preparation time $\Delta t < 1/ \lambda$ and $\Delta t < 1/ g$. The system initial conditions are $\left\langle  S_{ee}\right\rangle_{t=0} = 1$, $\left\langle  b^{\dagger 2}  b^2\right\rangle_{t=0} = 2\bar{n}^2$, $\left\langle  b^{\dagger}  b\right\rangle_{t=0} = \left\langle  S_{ee} b^{\dagger}  b \right\rangle_{t=0} = \bar{n}$, while all other variables are null at  	$t=0$. We consider a strong phonon-molecule interaction coupling $\lambda/\kappa_c=6$ when compared to the phonon frequency $\omega_{ph}/\kappa_c=2$. The thermal damping of the phonon mode is described by strong damping rates $\kappa/\kappa_c=1$ at low temperatures, i.e., $\bar{n}=1$. The system is truncated at $m_{max}=150$.

The behaviour of the mean phonon number $\left\langle  b^{\dagger} b\right\rangle$ of Fig.\ref{pop} when $s(t)=0$  is dictated by the phenomenon of phonon excitation due to the optomechanical interaction of the mechanical resonator with the molecular excited state, as well as, the damping phenomenon. Weaker optomechanical couplings $\lambda$ would generally lead to lower phonon excitation rates. However, the longer the molecule stays in its excited state the longer is the time interval when phonons are excited.
The estimated mean phonon number is also determined by the phonon damping and it will increase when decreasing the phonon losses, i.e., for lower damping rates $\kappa$.
Once the molecule decays to its ground state, the phonon dynamics are subject to thermal damping. The mechanical vibrations are damped to the environmental temperature of the thermal bath. This damping process will be slower for lower damping rates.

The frequency modulation of the molecular transition slows-down the molecular decay, as expected. 
In Fig.\ref{pop} we observe that long-lived phonons are obtained for a slow-downed population decay dynamics. The enhancement of the lifetime of the molecular excited state allows for phonons to be excited for a longer time period and, therefore, leads to an increase of the phonon effective lifetime.
The molecular decay dynamics is determined only by the optical setup and is independent of the phonon dynamics as expressed in eq.(\ref{eqs22}).
Therefore, long-lived phonons may be obtained for various optomechanical parameters.  

 %%%%%%%%%%%%%%%%%%%%%%%%%%%%%%%%%%%%%%%%%%%%%%%%%%%%%%%%%%%%%%
%%%%%%%%%%%%%%%%%%%%%%%%%%%%%%%%%%%%%%%%%%%%%%%%%%%%%%%%%%%%%%

The slow-down of the molecular decay affects the phonon distribution as well. In Fig.\ref{g2} we depict the temporal evolution of the second-order phonon-phonon correlation function $g^{(2)}(0)$ for the cases of modulated (blue solid line) and non-modulated (green dashed line) molecular transition frequency. The phonon field is initially in equilibrium with the thermal bath. Hence, $g^{(2)}(0)=2$ at $t=0$, which corresponds to an initial thermal distribution of the phonon field. The phonons become coherently distributed when $g^{(2)}(0)=1$. As depicted, long-lived phonons have a second-order correlation function near unity for longer time periods, corresponding to time intervals when most of the molecular population remains in excited state.
Therefore, the lifetime increase of the molecular excited state allows
the confinement of excited phonons in a quasi-coherent
state for a longer time.

The previously discussed behaviour of the phonon field can be obtained for different parameters, namely, lower damping rates $\kappa$ and lower optomechanical couplings $\lambda$. In such cases, a significant increase of phonon effective lifetimes occurs when the lifetime of the non-modulated molecular excited state is equal or longer than the corresponding phonon lifetime.
However, this results in lower values of the mean phonon number.  
This allows us to extrapolate the current discussion to other optomechanical setups based on other types of emitters such as quantum-dots, nitrogen vacancy in diamonds or superconducting circuits \cite{aspe14}. Moreover, the changes introduced by the transition frequency modulation to the molecular population decay dynamics are analytically expressed via the quasi-damping term $\gamma (t)$ of eqs.(\ref{eqs22} - \ref{eqg2}). There are various quantum models allowing the optical control of the behaviour of the spontaneous emission dynamics. The changes introduced to the decay dynamics of such systems may be analytically expressed via an effective spontaneous emission term \cite{lini08}.
Therefore, we suggest that similar behaviour in the phonon dynamics may be obtained in optomechanical devices using other methods of optical control of spontaneous emission.

Another field of application of the current investigation represents the molecular optomechanics with metal nanoparticles. Within this framework, the quantum dynamics of a metal nanoparticle resonantly interacting with a molecule may be expressed via the Hamiltonian of eq.(\ref{Hinit}) and the master equation of eq.(\ref{meq}), within the bad cavity limit \cite{hugh21}. In this scenario, the operators $a$ and $a^{\dagger}$ represent the plasmonic field of the nanoparticle, $b$ and $b^{\dagger}$ represent the molecular vibrations. The master equation should contain an additional dephasing term which, however, doesn't influence the obtained system of equations of motion of eqs.(\ref{eqs22} - \ref{eqg2}).

\section{Summary}
%%%%%%%%%%%%%%%%%%%%%%%%%%%%%%%%%%%%%%%%%%%%%%%%%

We have investigated the model of a two-level molecule with a slowed-down decay dynamics and embedded in a mechanical resonator made from an organic crystal. The slow-down of its spontaneous emission is achieved by modulating the molecular transition frequency and by placing the molecule in an optical cavity. As the molecular energetic states couple with the phonons of the host organic crystal, we have been able to demonstrate that the optical control of the molecular population dynamics, allows the control of the phonon dynamics. Namely, we have shown that a slow-down of the molecular decay dynamics enhances the effective lifetime of the vibrational quanta.
This effect becomes prominent for mechanical resonators with high damping rates, such as organic crystals.
Moreover, the obtained results may be extrapolated to other optomechanical setups based on other types of emitters, as well as, to molecular optomechanical models.

%%%%%%%%%%%%%%%%%%%%%%%%%%%%%%%%%%%%%%%%%%%%%%%%%%%%%%%%%%%%%%%%%%%

\section*{Acknowledgement}
%%%%%%%%%%%%%%%%%%%%%%%%%%%%%%%%%%%%%%%%%%%%%%%%%%%%%%%%%%%%%%%%%%%
We gratefully appreciate the financial support by the Moldavian National Agency for Research and Development, Grant No. 20.80009.5007.07.
%%%%%%%%%%%%%%%%%%%%%%%%%%%%%%%%%%%%%%%%%%%%%%%%%%%%%%%%%%%%%%%%%%

%%%%%%%%%%%%%%%%%%%%%%%%%%%%%%%%%%%%%%%%%%%%%%%%%%%%%%%%%%%%%%%%%%%

\section*{Disclosures}
%%%%%%%%%%%%%%%%%%%%%%%%%%%%%%%%%%%%%%%%%%%%%%%%%%%%%%%%%%%%%%%%%%%
The authors declare that there are no conflicts of interest related to this article.
%%%%%%%%%%%%%%%%%%%%%%%%%%%%%%%%%%%%%%%%%%%%%%%%%%%%%%%%%%%%%%%%%%

%%%%%%%%%%%%%%%%%%%%%%%%%%%%%%%%%%%%%%%%%%%%%%%%%%%%%%%%%%%%%%%%%%%

%%%%%%%%%%%%%%%%%%%%%%%%%%%%%%%%%%%%%%%%%%%%%%%%%%%%%%%%%%%%%%%%%%%%%%%%%%%
\end{document}